\documentclass[sigconf]{acmart}

\AtBeginDocument{%
  }

\setcopyright{acmlicensed}
\copyrightyear{2018}
\acmYear{2018}
\acmDOI{XXXXXXX.XXXXXXX}
\acmConference[Conference acronym 'XX]{Make sure to enter the correct
  conference title from your rights confirmation email}{June 03--05,
  2018}{Woodstock, NY}
\acmISBN{978-1-4503-XXXX-X/2018/06}

\usepackage{enumitem}
\usepackage{multirow}
\usepackage{makecell}

\begin{document}

\title{POEM: Partial-Order Enhanced Real-Time Sequential Modeling for Recommendation}

\author{Linxiao Che}
\authornote{These authors contributed equally to this work.} 
\email{chelinxiao@kuaishou.com}
\affiliation{
\institution{Kuaishou Technology}
\city{Beijing}
\country{China}
}

\author{Yijia Sun}
\authornotemark[1]
\email{sunyijia@kuaishou.com}
\orcid{0009-0000-8779-2108}
\affiliation{%
\institution{Kuaishou Technology}
\city{Beijing}
\country{China}
}

\author{Siyuan Lou}  
\email{lousiyuan05@kuaishou.com}
\affiliation{%
\institution{Kuaishou Technology}
\city{Beijing}
\country{China}
}

\author{Shanshan Huang}  
\email{huangshanshan@kuaishou.com}
\affiliation{%
\institution{Kuaishou Technology}
\city{Beijing}
\country{China}
}

\author{Qiang Luo}
\authornote{Corresponding authors.} 
\email{luoqiang@kuaishou.com}
\affiliation{
\institution{Kuaishou Technology}
\city{Beijing}
\country{China}
}

\author{Ruiming Tang}
\authornotemark[2]
\email{tangruiming@kuaishou.com}
\affiliation{%
\institution{Kuaishou Technology}
\city{Beijing}
\country{China}
}

\author{Han Li}
\authornotemark[2]
\email{lihan08@kuaishou.com}
\affiliation{%
\institution{Kuaishou Technology}
\city{Beijing}
\country{China}
}

\author{Kun Gai}
\email{gai.kun@qq.com}
\affiliation{%
\institution{Unaffiliated}
\city{Beijing}
\country{China}
}

\renewcommand{\shortauthors}{Linxiao Che et al.}

\begin{abstract}
Real-time recommendation systems face the challenge of dynamically changing user interests and contextual environments. Traditional sequential recommendation models rely on static historical click sequences, which struggle to capture real-time preference shifts and ignore the structured information embedded in the system's internal ranking logic. This paper proposes \textbf{POEM} (\textbf{P}artial-\textbf{O}rder \textbf{E}nhanced \textbf{M}odeling), a novel real-time sequential modeling framework that leverages the \textbf{partial-order relations} inherent in the recommendation cascade. POEM innovatively utilizes the real-time multi-task ranking scores (e.g., predicted CTR, watch time) from the preceding ranking stage as supervisory signals to construct dynamic partial-order sequences, thereby achieving fine-grained real-time interest modeling and aligning system objectives with user behavior. Our contributions are threefold: a \textbf{partial-order guided sequence construction paradigm} that augments traditional temporal sequences with a dynamically grouped and sampled sequence based on real-time ranking scores, enabling per-request interest reassessment; (2) a \textbf{multi-objective score fusion mechanism} that integrates various ranking signals through normalized rank-weighting into a unified quintuple representation; and (3) a \textbf{hierarchical sample learning strategy} combining system-preferred items (top-ranked) and user feedback (e.g., long-play videos) as positive samples, enhanced by graph-retrieved hard negative samples and a margin-based pairwise loss.
Deployed in Kuaishou's platform, POEM achieves significant online gains: \textbf{+0.249\%} and \textbf{+0.213\%} in average viewing time per user on the KS Single Page and KS Lite Page, respectively. Extensive ablation studies validate the effectiveness of each component, demonstrating POEM's superiority in real-time responsiveness, recommendation accuracy, and content diversity.
\end{abstract}




\maketitle

\section{Introduction}
\label{sec:introduction}

The pursuit of real-time, precise user interest modeling lies at the heart of modern information-rich platforms, such as short-video feeds and e-commerce. In these dynamic environments, a user's intent can shift rapidly, influenced by immediate context and the presented content itself. Sequential recommendation models have become a cornerstone technique by modeling user behavior over time. Models like SASRec \cite{kang2018self} and BERT4Rec \cite{sun2019bert4rec}, often built upon Transformer architectures, excel at capturing sequential dependencies within a user's \textbf{historical click sequence}.

However, a fundamental disconnect exists between these models and the industrial recommendation pipeline. These models typically operate on a \textbf{static snapshot} of past clicks, treating the recommendation system as a mere observer. In reality, the system is an active participant: its intricate ranking stages continuously generate rich, multi-faceted estimations of user preference—such as predicted click-through rate (CTR), conversion rate (CVR), and watch time—for thousands of candidate items in real-time. These \textbf{ranking signals} form a \emph{partial order} that reflects the system's current, context-aware understanding of what the user might prefer next. This valuable source of supervisory signal remains largely untapped by traditional sequential recommenders, which rely solely on the \emph{observed} user clicks, a signal that is sparse, delayed, and lacks the granular preference discrimination inherent in ranking scores.

This oversight leads to two key limitations: \textbf{(1) Underutilization of System Intelligence:} Valuable preference information from the ranking cascade is ignored, limiting model expressiveness. \textbf{(2) Inherent Latency in Real-Time Modeling:} Models depend on new user clicks to update their sequence view. In a fast-paced feed, a user's interest might evolve based on unseen items or changing context long before a new click occurs, leaving the model with a stale representation.

To bridge this gap, we propose \textbf{POEM (Partial-Order Enhanced Modeling)}, a novel framework that rethinks sequence construction for real-time recommendation. The core thesis of POEM is: \emph{the real-time ranking scores from the system can serve as a powerful supervisory signal to \textbf{construct} the input sequence itself, guiding the model towards learning a representation that mirrors the system's current, preference-aware view of the user.}

In POEM, for each user request, we dynamically construct a sequence not only from the user's past clicks, but also from the \textbf{top-ranked candidate items of the user's previous refresh request}. These items are sorted and grouped by their fused multi-objective ranking scores, and then randomly sampled within groups. This yields a \emph{partial-order structured sequence} that encapsulates the system's preference hierarchy. A special $[CLS]$ token aggregates information from this sequence to form the user's instantaneous interest embedding for retrieval. This approach ensures the user representation is \textbf{re-evaluated on every request} based on the freshest contextual signals, achieving true request-level real-time modeling.

To effectively learn from this novel sequence paradigm, we design a complementary hierarchical learning strategy. We define positive samples as a combination of system-predicted top-ranked items and actual user feedback (e.g., long-play videos), aligning the learning objective with both system efficiency and user satisfaction. Furthermore, we employ a dynamic, graph-based strategy to mine hard negative samples tailored to the current context, enhancing the model's discriminative power for fine-grained partial-order learning.

We evaluate POEM comprehensively on large-scale industrial data from Kuaishou. 
Online A/B tests demonstrate significant improvements across two major scenarios: \textbf{+0.249\% in average usage time per user} on the KS Single Page and \textbf{+0.213\%} on the KS Lite Page (see Table~\ref{tab:online_ab}).
Rigorous offline ablation studies confirm the individual contribution of each component: the partial-order construction paradigm, multi-signal fusion, and the hierarchical learning strategy.

Our main contributions are summarized as follows:
\begin{itemize}
    \item We propose a novel \textbf{partial-order guided sequence construction framework} for real-time recommendation. It dynamically builds user interest sequences from real-time multi-objective ranking scores, replacing the conventional reliance on static chronological click histories.
    
    \item Within this framework, we introduce a \textbf{multi-objective signal fusion mechanism} to unify heterogeneous ranking signals, and design a \textbf{hierarchical learning objective} that aligns system predictions with genuine user feedback, supported by context-aware hard negative sampling.
    
    \item We implement the proposed POEM framework end-to-end in a large-scale industrial platform and validate its effectiveness through rigorous offline analyses and online A/B tests, demonstrating significant gains in both recommendation accuracy and real-time responsiveness.
\end{itemize}

The rest of this paper is organized as follows. We review related work in Section~\ref{sec:related_work}. The POEM framework is detailed in Section~\ref{sec:method}. Experimental setup and results are presented in Section~\ref{sec:experiments}, along with case study. We finally go to conclusion in Section~\ref{sec:conclusion}.

\begin{table*}[htbp]
\centering
\caption{Positioning of the proposed POEM framework against related paradigms.}
\label{tab:positioning}
\begin{tabular}{p{4cm} p{4cm} p{4cm} p{4cm}}
\toprule
\textbf{Category} & \textbf{Sequence Input} & \textbf{Use of Ranking Signals} & \textbf{Update Granularity} \\
\midrule
Classical Sequential Models & Static chronological interaction history & Not utilized & New user interaction \\

LTR-informed Retrieval & Static chronological interaction history & As an auxiliary loss or supervision & Model re-training / fine-tuning \\
\textbf{POEM (Ours)} & \textbf{Both interatcion history and dynamic partial-order sequence} & \textbf{To construct the input sequence} & \textbf{Every user request} \\
\bottomrule
\end{tabular}
\end{table*}

\section{Related Work}
\label{sec:related_work}

\subsection{Sequential Recommendation}
\label{subsec:sequential_rec}
Sequential recommendation focuses on predicting a user's next interaction by capturing patterns in their behavior history. Early approaches were based on Markov Chains and RNNs \cite{rendle2010factorizing, hidasi2015session, tang2018caser}. The introduction of the Transformer architecture \cite{vaswani2017attention} revolutionized the field, leading to models like SASRec \cite{kang2018self} and BERT4Rec \cite{sun2019bert4rec} that model long-range dependencies. Recent advances include CL4SRec \cite{2022cl4srec} which incorporates contrastive learning, and MPFormer \cite{sun2025mpformer} which introduces a multi-task Transformer for heterogeneous patterns. Mao et al. \cite{mao2024analysis} propose a dynamic user interest model to capture high-frequency preference shifts.

A core challenge remains the modeling of \textit{evolving user interests}. Methods like DIN \cite{zhou2018deep}, DIEN \cite{zhou2019deep}, and MIMN \cite{lu2025mimn} design specialized architectures to capture interest shifts over time, while SIM \cite{pi2020search} retrieves relevant long-term behaviors. However, even these models operate on a static, chronologically ordered sequence of historical interactions, and their updates are triggered only by new observed feedback. In contrast, POEM \textit{dynamically constructs} a sequence from real-time ranking scores, allowing the user representation to refresh on every request, even in the absence of new interactions.

\subsection{Context-Aware Recommendation}
\label{subsec:context_aware}

Contextual information—such as time, location, and device—plays a crucial role in personalizing recommendations. A wide range of techniques have been developed to integrate such signals. Shallow models like Factorization Machines \cite{koren2009matrix} efficiently model feature interactions. Deep learning-based architectures, including DeepFM \cite{guo2017deepfm} and xDeepFM \cite{lian2018xdeepfm}, further enhance the capacity to learn high-order feature combinations. In sequential settings, context can be integrated through time-aware attention mechanisms \cite{li2020time} or by attending to contextually relevant historical behaviors \cite{guo2025context}. These methods typically treat context as auxiliary features that are fused with user and item representations.

Cutting-edge research now focuses on sequence refinement. MGSD-WSS \cite{liu2025mgsd} utilizes weakly supervised signals to denoise historical sequences, ensuring that only contextually relevant behaviors are preserved. However, these methods typically treat context as auxiliary features to be fused with representations. POEM leverages \textit{real-time multi-objective ranking scores} as a more fundamental, intent-reflective context, using them to structurally guide the construction of the input sequence itself.

\subsection{Utilizing Ranking Signals in Recommendation} \label{subsec:ranking_signals}

Ranking signals, which have long been pivotal in the final re-ranking stage of recommender systems \cite{burges2010ranknet, ma2018entire, pang2020setrank}, have recently been explored as valuable feedback to enhance earlier stages such as candidate retrieval. One line of work leverages these signals for direct supervision, such as using sampling-bias correction \cite{yi2019sampling} or adapting listwise losses \cite{bruch2020stochastic, wilm2023scaling} to better align retrieval models with ranking objectives. Another common approach employs knowledge distillation, where a complex ranker's knowledge is transferred to a lightweight retrieval model to bridge the performance gap between stages \cite{tang2018ranking, khani2024bridging}. More recently, research has focused on joint or iterative optimization frameworks that seek deeper integration\cite{sun2025grank}. These efforts aim to resolve the systemic inconsistency between stages, either through algorithmic refinements like iterative ensemble sorting \cite{cao2025pantheon} or reasoning-chain guided generation \cite{wu2025rankcot}, or through architectural designs that explicitly incorporate real-time bids signals \cite{liu2025bidding} or optimize inter-stage consistency \cite{zhao2023copr}.

A key distinction of our work lies in how ranking signals are structurally utilized. Unlike prior paradigms that primarily treat ranking scores as a downstream \textit{supervision signal} \cite{yi2019sampling, cao2025pantheon} or a source for \textit{distillation} \cite{kang2023distillation, khani2024bridging}, POEM repurposes the \textit{partial order} implied by real-time ranking scores as a structural prior. This order is used to dynamically assemble the model's input sequence, thereby shifting the role of ranking signals from an optimization target to a generative guide for representation learning.

\subsection{Positioning Summary}

Table~\ref{tab:positioning} summarizes the key differences between POEM and related paradigms. POEM introduces a novel framework that bridges real-time system intelligence (via ranking signals) and sequential modeling by dynamically constructing sequences based on a preference-derived partial order. This enables a unique form of request-level real-time user interest modeling for the recall stage.

\section{Methodology}
\label{sec:method}

In this section, we present the POEM framework in detail. We first provide an overview of the entire system and then elaborate on each key component: the dynamic partial-order sequence construction, multi-objective signal fusion, hierarchical learning objective, and the real-time serving architecture.

\subsection{Overall Framework}
\label{subsec:framework}

\begin{figure*}[t]
    \centering
    \includegraphics[width=0.9\linewidth]{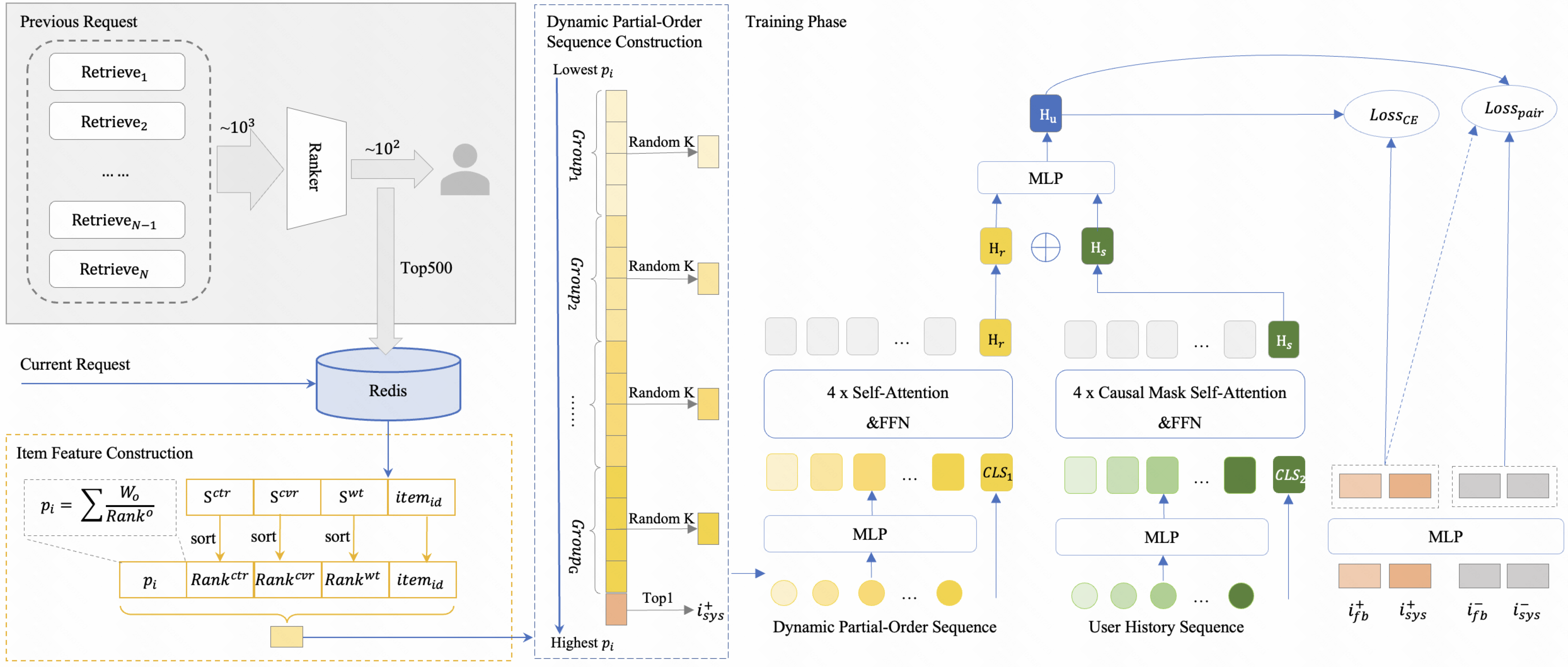}
    \caption{The POEM framework workflow. 
    \textbf{Left (Sequence Construction):} For a current request, we retrieve the top-ranked items from the user's \textit{previous request} (stored in Redis). The \textit{Item Feature Construction} module calculates a priority score $p_i$ by aggregating multi-dimensional signals ($S^{ctr}, S^{cvr}$, etc.). These items are then partitioned into $G$ groups based on $p_i$, from which a \textit{Dynamic Partial-Order Sequence} is formed via random sampling within each group.
    \textbf{Right(Training Phase):} The model uses dual encoders to process the partial-order sequence and user history, generating the representation $H_u$ via an MLP. The training is optimized by a hierarchical loss combining cross-entropy and pairwise objectives ($Loss_{CE}$ and $Loss_{pair}$) that considers both user feedback ($i_{fb}$) and system signals ($i_{sys}$).}
    \label{fig:framework}
    \vspace{-0.2in} %
\end{figure*}

POEM aims to learn a real-time user interest representation $\mathbf{h}_u \in \mathbb{R}^d$ that reflects the system's current understanding of user preferences based on real-time ranking signals. As illustrated in Figure~\ref{fig:framework}, POEM operates in two phases: training and serving.

During training, for each user request instance, POEM processes two complementary sequences: (1) a historical interaction sequence $S_h$ constructed from the user's past interactions, and (2) a dynamic partial-order sequence $S_r$ constructed from the top-ranked candidate items of the \textbf{same user's previous refresh request}. These sequences are encoded by independent Transformers, and their fused representations produce $\mathbf{h}_u$, which is optimized via a hierarchical learning objective.

During serving, for each new user request, POEM constructs $S_r$ from the top-ranked candidate items of the \textbf{same user's previous refresh request}, processes it alongside $S_h$, and generates $\mathbf{h}_u$ for real-time ANN-based retrieval. This design ensures temporal causality and leverages the most recent user interest signals.

\subsection{Dynamic Partial-Order Sequence Construction}
\label{subsec:sequence_construction}
The dynamic partial-order sequence $S_r$ is the core innovation of POEM, capturing the system's real-time understanding of user preferences. It is constructed from the top-ranked candidate items of the same user's previous refresh request, ensuring temporal causality between training and serving. Specifically, let $\mathcal{C} = \{i_1, i_2, ..., i_N\}$ denote the set of candidate items from the previous refresh request. For each item $i$, we have multiple real-time ranking scores from the preceding ranking stage, such as predicted CTR ($s^{\text{ctr}}$), CVR ($s^{\text{cvr}}$), and watch time ($s^{\text{wt}}$). We denote the vector of ranking scores for item $i$ as $\mathbf{s}_i = [s^{\text{ctr}}_i, s^{\text{cvr}}_i, s^{\text{wt}}_i]$.

\textbf{Score Fusion and Ranking.}
To establish a consistent partial ordering, we compute a fused score \(p_i\) for each item. Direct averaging of per-objective scores is suboptimal due to scale misalignment and varying objective importance. We therefore propose a fusion scheme based on ranking and weighting.

For each objective \(o \in \{\textrm{ctr}, \textrm{cvr}, \textrm{wt}\}\), we rank all items in the candidate set \(\mathcal{C}\) by their raw scores \(s^o\) in descending order, with rank 1 being the highest. Let \(\operatorname{rank}^o(i)\) denote the rank of item \(i\) under objective \(o\). We define a per-objective utility score as the inverse rank:
\begin{equation}
    s^o(i) = \frac{1}{\operatorname{rank}^o(i)}.
\end{equation}
Since the candidate set size is fixed, \(s^o(i)\) shares the same numerical range across objectives, making additional normalization unnecessary.

The final fused score \(p_i\) is then computed as a weighted combination:
\begin{equation}
    p_i = \sum_{o \in \{\textrm{ctr}, \textrm{cvr}, \textrm{wt}\}} w^o \cdot s^o(i),
\end{equation}
where \(w^o\) is the weight for objective \(o\) and item \(i\). The candidate set \(\mathcal{C}\) is sorted in ascending order of \(p_i\) to produce the final ranking.

\textbf{Grouping and Sampling.} To construct a sequence that reflects the partial order while maintaining diversity and controlling length, we adopt a grouping and sampling strategy. The sorted list is divided into $G$ contiguous groups of approximately equal size. From each group, we randomly sample $K$ items without replacement. The sampled items from group 1 to $G$ are concatenated to form the sequence $S = [i_{(1,1)}, ..., i_{(1,K)}, i_{(2,1)}, ..., i_{(G,K)}]$. This approach ensures that the sequence broadly respects the global order (items from higher-score groups appear earlier) while introducing stochasticity within each group. In our implementation, we set $G=8$ and $K=8$, resulting in a sequence length of 64.

\textbf{Sequence Representation.} Each item $i$ in $S$ is represented as a quintuple:
\begin{equation}
\mathbf{r}_i = \left( \text{rank}^{\text{ctr}}(i), \text{rank}^{\text{cvr}}(i), \text{rank}^{\text{wt}}(i), p_i, \text{ID}(i) \right),
\end{equation}
where the $\text{ID}(i)$ is the item's unique identifier. To obtain a dense embedding \(\mathbf{e}_i \in \mathbb{R}^d\), the quintuple is projected through a hybrid encoding layer:
\[
\mathbf{e}_i = \mathbf{W}_{\text{feat}} \cdot \left[ \operatorname{rank}^{\text{ctr}}(i),\; \operatorname{rank}^{\text{cvr}}(i),\; \operatorname{rank}^{\text{wt}}(i),\; p_i \right]^{\top} + \mathbf{E}_{\text{embed}}\bigl(\operatorname{ID}(i)\bigr),
\]
where \(\mathbf{W}_{\text{feat}} \in \mathbb{R}^{d \times 4}\) is a learnable weight matrix for the continuous features, and \(\mathbf{E}_{\text{embed}}\) denotes the item embedding table. The sequence of all embedded items \(\mathbf{S_r} = [\mathbf{e}_1, \mathbf{e}_2, \dots, \mathbf{e}_{L}]\) (with \(L = G \times K = 64\)) forms the input to the subsequent transformer encoder.

\subsection{User Representation Learning via Dual Sequences}
\label{subsec:user_representation}

POEM processes two complementary sequences to learn a comprehensive user representation:

\textbf{Historical Interaction Sequence ($S_h$).} We construct a fixed-length sequence of the user's recent interactions for modeling long-term preferences. Each interaction is represented by concatenating multiple feature embeddings including item ID, content tags, and auxiliary signals.

\textbf{Dual-Sequence Encoding.} We employ two independent Transformer encoders \cite{vaswani2017attention} to process the two sequences in parallel. Let $\text{Encoder}_h$ denote the encoder for the historical interaction sequence $S_h$, and $\text{Encoder}_r$ denote the encoder for the dynamic partial-order sequence $S_r$. Both encoders follow the standard Transformer architecture with multiple layers of multi-head self-attention and position-wise feed-forward networks, sharing the same architectural configuration but not parameters.

For each encoder, we append a special token to the end of its input sequence to aggregate sequence-level information. Let $\mathbf{h}_h \in \mathbb{R}^d$ and $\mathbf{h}_r \in \mathbb{R}^d$ denote the output vectors corresponding to these special tokens from the two encoders, respectively. We then fuse these two representations through concatenation followed by a multi-layer perceptron (MLP) to obtain the final user representation:

\begin{equation}
\mathbf{h}_u = \text{MLP}\left([\mathbf{h}_h; \mathbf{h}_r]\right),
\end{equation}

where $[\cdot;\cdot]$ denotes vector concatenation, and the MLP consists of two linear layers with a non-linear activation in between.

\begin{table*}[thbp]
\centering
\caption{Offline performance comparison of POEM and baseline models on test sets of different temporal granularities (1-hour and 1-day). Best results are in \textbf{bold}, second-best are \underline{underlined}. Relative gains for POEM are computed against the second-best baseline in each setting.}
\label{tab:offline_results}
\begin{tabular}{lcccccccc}
\toprule
& \multicolumn{4}{c}{\textbf{1-hour Test Set}} & \multicolumn{4}{c}{\textbf{1-day Test Set}} \\
\cmidrule(lr){2-5} \cmidrule(lr){6-9}
\textbf{Model} & \textbf{HR@50} & \textbf{NDCG@50} & \textbf{HR@100} & \textbf{NDCG@100} & \textbf{HR@50} & \textbf{NDCG@50} & \textbf{HR@100} & \textbf{NDCG@100} \\
\midrule
DSSM & 0.4703 & 0.1778 & 0.6070 & 0.1999 & 0.3951  & 0.1484 & 0.5132 & 0.1676 \\
SASRec & 0.5267 & 0.2092 & 0.6568 & 0.2303 & 0.4419 & 0.1746 & 0.5547 & 0.1929 \\
CAIN variant & \underline{0.5749} & \underline{0.2371} & \underline{0.7008} & \underline{0.2575} & \underline{0.4700} & \underline{0.1923} & \underline{0.5791} & \underline{0.2100} \\
\textbf{POEM} & \textbf{0.6661} & \textbf{0.3080} & \textbf{0.7725} & \textbf{0.3253} & \textbf{0.4879} & \textbf{0.2100} & \textbf{0.5833} & \textbf{0.2255} \\
\midrule
\makecell{Relative Gain} & +15.9\% & +29.9\% & +10.2\% & +26.3\% & +3.81\% & +9.19\% & +0.72\% & +7.37\% \\
\bottomrule
\end{tabular}
\end{table*}

\subsection{Hierarchical Learning Objective}
\label{subsec:learning_objective}

We design a learning objective that aligns the user representation $\mathbf{h}_u$ with items that reflect both system preference and genuine user satisfaction.

\textbf{Positive Samples.} For each training instance, we define two types of positive items to bridge the gap between user behavior and system intelligence:
\begin{itemize}
    \item \textbf{User-Feedback Positive ($i^+_{\text{fb}}$)}: An item from $\mathcal{C}$ that was actually exposed to the user and received positive feedback. This grounds the learning in actual user behavior.
    \item \textbf{System-Preferred Positive ($i^+_{\text{sys}}$)}: The item with the highest multi-objective score $p_i$ among the candidate set $\mathcal{C}$. This represents the item the system believes the user is most likely to engage with positively.
\end{itemize}

While $i^+_{\text{fb}}$ represents the ultimate goal, it is often sparse and subject to selection bias. In contrast, $i^+_{\text{sys}}$ serves as a \textit{distillation signal} from the sophisticated ranking stage. By incorporating $i^+_{\text{sys}}$, POEM ensures that the retrieval model is \textbf{cascade-aware}, generating candidates that are more likely to be prioritized by downstream rankers, thereby improving the overall delivery rate in the production pipeline.
The final positive set for the instance is $\mathcal{P} = \{i^+_{\text{sys}}, i^+_{\text{fb}}\}$.

\textbf{Hard Negative Mining via Dynamic Graph.} In-batch random negative sampling often yields items that are too easy to distinguish. To enhance discriminability, we mine hard negatives from a global item-item similarity graph $\mathcal{G}$ constructed offline using the Swing algorithm \cite{yang2020large}, which captures co-occurrence patterns. Specifically, for each positive item $i^+ \in \mathcal{P} = \{ i^+_{fb}, i^+_{sys} \}$, we independently retrieve its top-$M$ most similar neighbors in $\mathcal{G}$, denoted as $\mathcal{N}(i^+, M)$. This process results in two positive-conditioned item-to-item negative pairs: the user-feedback negative $i^-_{fb}$ mined around $i^+_{fb}$, and the system-preferred negative $i^-_{sys}$ mined around $i^+_{sys}$, respectively, ensuring consistency with the hierarchical training structure illustrated in Figure~\ref{fig:framework}. To balance hardness and diversity, we adopt a mixed strategy: with probability $\rho$, we select the item with the \textit{third-lowest} similarity score within $\mathcal{N}(i^+, M)$ as the hard negative $i^-_{\text{hard}}$; otherwise, we randomly sample an item from $\mathcal{N}(i^+, M)$. This strategy, informed by empirical analysis, provides negatives that are semantically related but likely not preferred in the current context. The mining is performed dynamically for each training instance, ensuring that both $(i^+_{fb}, i^-_{fb})$ and $(i^+_{sys}, i^-_{sys})$ pairs remain aligned with the specific partial-order context.

\textbf{Loss Function.} We combine a standard cross-entropy loss with a margin-based pairwise loss to form the hierarchical objective. 
\begin{equation}
\mathcal{L} = \mathcal{L}_{\text{CE}} + \lambda \mathcal{L}_{\text{pair}},
\end{equation}

where $\lambda$ is a balancing hyperparameter.

The cross-entropy loss $\mathcal{L}_{\text{CE}}$ uses in-batch negatives:

\begin{equation}
\mathcal{L}_{\text{CE}} = -\frac{1}{B} \sum_{b=1}^{B} \log \frac{\exp(\text{sim}(\mathbf{h}_u^b, \mathbf{e}_{i_b^+}) / \tau)}{\sum_{j \in \mathcal{N}_b} \exp(\text{sim}(\mathbf{h}_u^b, \mathbf{e}_j) / \tau)},
\end{equation}
where $\mathbf{h}_u^b$ is the user representation for the $b$-th instance, $i_b^+$ is a positive item sampled from $\mathcal{P}_b$, $\mathcal{N}_b$ is the set of negative items (in-batch negatives) for this instance, $\text{sim}(\cdot,\cdot)$ denotes cosine similarity, and $\tau$ is a temperature parameter. Additionally, we incorporate a sampling bias correction mechanism \cite{yi2019sampling} to mitigate popularity bias in training data.

The pairwise loss $\mathcal{L}_{\text{pair}}$ explicitly enforces a margin between the positive and hard negative similarities:

\begin{equation}
\mathcal{L}_{\text{pair}} = \frac{1}{B} \sum_{b=1}^{B} \max\left(0, \gamma + \text{sim}(\mathbf{h}_u^b, \mathbf{e}_{i^-_{\text{hard},b}}) - \text{sim}(\mathbf{h}_u^b, \mathbf{e}_{i_b^+})\right),
\end{equation}
where $\gamma > 0$ is the margin. This loss directly optimizes the model to distinguish between the positive item and a challenging alternative.

\subsection{Real-Time Serving Architecture}
\label{subsec:serving}

To deploy POEM online, we design a low-latency serving pipeline that strictly adheres to the temporal causality principle: the user representation for the \textit{current} request is derived from ranking signals of the \textit{previous} request. The real-time process for each user request is as follows:

\begin{enumerate}
    \item \textbf{Input Signal Preparation:} For the incoming request, the system fetches two key inputs: (a) the user's historical interaction sequence $S_h$; and (b) the list of top-ranked items and their multi-objective ranking scores $\{\mathbf{s}_i\}$ from the user's \textbf{immediately preceding request}, which are retrieved from a low-latency cache (e.g., Redis).

    \item \textbf{Sequence Encoding \& Inference:} The dynamic sequence $S_r$ is constructed from previous request signals (Section~\ref{subsec:sequence_construction}) and encoded together with $S_h$ by the POEM model to produce $\mathbf{h}_u$. This step is deeply optimized for millisecond-level latency.

    \item \textbf{ANN Retrieval:} $\mathbf{h}_u$ is used as a query to retrieve the final candidate set from a pre-built Approximate Nearest Neighbor (ANN) index over the entire item corpus.

    \item \textbf{Feedback Loop for Subsequent Requests:} Concurrently, the candidate items for the \textit{current} request proceed through downstream ranking stages (e.g., coarse and fine ranking). The final top-ranked items \textbf{from all recall sources} along with their scores are cached. This cached data then serves as the input for constructing $S_r$ when the same user makes the \textit{next} refresh request, thereby closing the real-time adaptation loop.
\end{enumerate}

This serving architecture ensures that every user request is served with an embedding that incorporates the system's most recent understanding of user preference, which is precisely one request cycle fresh. It fulfills the design goal of request-level real-time interest modeling while maintaining the low latency required for large-scale production.

\section{Experiments}
\label{sec:experiments}

We conduct comprehensive offline and online experiments to evaluate the effectiveness of the proposed POEM framework. The experiments are designed to answer the following research questions:
\begin{itemize}
    \item \textbf{RQ1}: Does POEM outperform state-of-the-art sequential recommendation baselines, both offline and online?
    \item \textbf{RQ2}: How does each component of POEM (dynamic sequence construction, multi-signal fusion, and hierarchical sample strategy) contribute to its performance?
    \item \textbf{RQ3}: Is POEM effective in improving recommendation diversity and real-time responsiveness?
\end{itemize}

\subsection{Experimental Setup}
\label{subsec:setup}

\subsubsection{Dataset}
We use a large-scale industrial dataset collected from a global-scale short-video recommendation platform serving over 400 million daily active users with more than 50 billion daily interactions. 
To ensure a robust offline evaluation, we construct a large-scale benchmark from six consecutive days of production logs.We split the data chronologically: the first five days for training, and the first hour of the sixth day for testing, to better match the real-time characteristics of industrial recommendation systems

Each training/testing instance corresponds to a user request. The candidate item set (approximately 1000 items) for this request is constructed from the \textbf{top-ranked results of the user's previous request}, with the user's true feedback (e.g., clicks, long-play) on these items serving as the label.

\begin{table}[h]
\centering
\small
\renewcommand{\arraystretch}{0.85}
\caption{Statistics of the experimental dataset.}
\label{tab:dataset_stats}
\begin{tabular}{@{}lrlr@{}}
\toprule
\multicolumn{2}{c}{\textbf{Left Group}} & \multicolumn{2}{c}{\textbf{Right Group}} \\
\cmidrule(r){1-2} \cmidrule(l){3-4}
\textbf{Metric} & \textbf{Value} & \textbf{Metric} & \textbf{Value} \\
\midrule
\# Users & 108M & \# Training Samples & 2.7B \\
\# Items & 42M & \# 1-hour Test & 13M  \\
Avg. Seq. Length & 64 & \# 1-day Test & 457M\\
Avg. Pos. $i^+_{\text{fb}}$ & 1.2 & & \\
\bottomrule
\end{tabular}
\end{table}

\subsubsection{Baseline Models}
We compare POEM with the following representative baseline models:
\begin{itemize}[leftmargin=*]

    \item \textbf{DSSM} \cite{huang2013dssm}: A classic non-sequential dual-tower model that learns static semantic embeddings. It provides a baseline to contrast static matching with POEM's use of dynamic ranking signals.
    
    \item \textbf{SASRec} \cite{kang2018self}: A unidirectional Transformer-based sequential model with self-attention.
    
    \item \textbf{CAIN (variant)} \cite{guo2025context}: Originally a ranking model leveraging Temporal Convolutional Networks (TCN) for local contextual modeling, we adapt it for the retrieval setting by retaining only its TCN-based user encoder. This serves as a strong context-aware baseline, contrasting its local sequence modeling with POEM's partial-order-driven representation learning.

\end{itemize}

\subsubsection{Evaluation Metrics}
We adopt both offline and online evaluation metrics.
\textbf{Offline metrics} computed on the test set include:
\begin{itemize}[leftmargin=*]
    \item \textbf{Hit Rate (HR@K)}: Proportion of test positive items appearing in the top-K recommendation list.
    \item \textbf{Normalized Discounted Cumulative Gain (NDCG@K)}: Ranking quality metric.
\end{itemize}

\subsubsection{Implementation Details}
All models are implemented in Tensorflow and trained in identical hardware/software environments. We use the AdamW optimizer with an initial learning rate of 0.001 and cosine annealing scheduler. The batch size is set to 1024. For POEM, the Transformer encoder has 4 layers, hidden dimension 128, and 1 attention heads. The sequence construction parameters are: number of groups $G=8$, samples per group $K=8$, sequence length 64. The loss balancing weight $\lambda=0.1$ and margin $\gamma=0.2$.

\subsection{Main Results(RQ1\&RQ3)}
\label{subsec:main_results}

\subsubsection{Offline Results}
Table~\ref{tab:offline_results} presents the offline performance comparison on a \textbf{1-hour test set} collected immediately after the training period. POEM significantly outperforms all baseline models across all metrics. Specifically, compared to the strongest sequential baseline CAIN Variant, POEM achieves a relative improvement of 15.9\% in HR@50 and 29.9\% in NDCG@50. 

\subsubsection{Analysis: Temporal Generalization of POEM}

To investigate POEM's real-time adaptability, we contrast its performance on the \textbf{1-hour} and \textbf{1-day} test sets (Table~\ref{tab:offline_results}). While POEM maintains a consistent lead over all baselines in both settings, its relative advantage is markedly more pronounced on the 1-hour set. This stems from the nature of $S_r$: the partial-order sequence is constructed from top-ranked candidates of the \textit{preceding} request, making it highly sensitive to real-time shifts in user interest, content pool, and ranking policy. Training and evaluation on a short, contiguous window allows POEM to closely track the current system state, whereas day-level evaluation spans a longer horizon over which the underlying recommendation logic and user distribution may shift, diluting signal precision.

To further quantify this temporal sensitivity, we partition test requests by time gap from the training cutoff and report HR@50 for POEM and the strongest sequential baseline (CAIN Variant) in Table~\ref{tab:fine_temporal}.

\begin{table}[htbp]
\centering
\caption{Fine-grained temporal performance (HR@50) of POEM vs. CAIN Variant over different time gaps.}
\begin{tabular}{lccc}
\toprule
Interval & POEM & CAIN Variant & Relative gain \\
\midrule
{[}1h, 2h)   & 0.6661 & 0.5749 & +15.9\% \\
{[}2h, 4h)   & 0.6259 & 0.5757 & +8.72\% \\
{[}4h, 8h)   & 0.5732 & 0.5239 & +9.35\% \\
{[}8h, 12h)  & 0.5376 & 0.4999 & +7.55\% \\
{[}12h, 24h) & 0.4801 & 0.4694 & +2.29\% \\
\bottomrule
\end{tabular}
\label{tab:fine_temporal}
\end{table}

As the time gap increases, the gains generally decrease—a trend that aligns with the expectation that cached ranking signals become less aligned with the current ranking policy and user interest distribution. A small fluctuation between the [2h,4h) and [4h,8h) buckets is observed, which we attribute to statistical noise. This fine-grained analysis complements the temporal generalization discussion above and provides direct evidence that POEM's real-time adaptation capability is a key driver of its superior performance, especially within short latency windows.

\subsubsection{Online A/B Test Results}
We deployed POEM on a leading short video platform for online recommendation services and conducted a 7-day online A/B test. The treatment group (5\% traffic) used POEM, while the control group (5\% traffic) used the production model Kuaiformer. Table~\ref{tab:online_ab} summarizes the core business metrics. POEM brings significant improvements: usage time per user increased by 0.249\% and 0.213\% on different pages. All improvements are statistically significant.

\begin{table}[htbp]
\centering
\caption{Online A/B Test Results of POEM Across Different Scenarios}
\label{tab:online_ab}
\begin{tabular}{lccccccc}
\toprule
Applications & \makecell{Total App \\ Usage Time} & \makecell{Usage Time \\ per User} & \makecell{Video \\ Watch Time} \\
\midrule
KS Single Page  & +0.254\% & +0.249\% & +0.237\% \\
KS Lite Page  & +0.158\% & +0.213\% & +0.335\% \\
\bottomrule
\end{tabular}
\end{table}

\subsection{Ablation Studies(RQ2)}
\label{subsec:ablation}

We conduct systematic ablation studies to validate the contribution of each core component of POEM. All ablations are performed on the same dataset with identical hyperparameter settings.

\subsubsection{Impact of Sequence Composition}
To verify the necessity of incorporating dynamic ranking signals, we compare POEM with variants using different input sequence compositions:
\begin{itemize}
    \item \textbf{Hist-Only}: Only uses the user's historical interaction sequence (e.g., the last 64 clicked videos), similar to traditional sequential models like SASRec.
    \item \textbf{RankSig-Only}: Uses only the dynamic partial-order sequence $S_r$, constructed from the fused ranking scores ($p_{partial}$) of the user's previous request, without incorporating historical interactions.
    \item \textbf{POEM (Hybrid)}: The full model combining both historical intent and real-time candidate-aware signals.
\end{itemize}

\begin{table}[htbp]
\centering
\caption{Ablation study on input sequence composition. }
\label{tab:ablation_sequence}
\begin{tabular}{lcccc}
\toprule
\textbf{Variant} & \textbf{\makecell{HR\\@50}} & \textbf{\makecell{NDCG\\@50}} & \textbf{\makecell{HR\\@100}} & \textbf{\makecell{NDCG\\@100}} \\
\midrule
Hist-Only & 0.5808  & 0.2531 & 0.6971 & 0.2720 \\
RankSig-Only  & 0.6511  & \textbf{0.3238} & 0.7456 & \textbf{0.3391} \\
\midrule
\textbf{POEM} & \textbf{0.6661}  & 0.3079 & \textbf{0.7725} & 0.3253 \\
\bottomrule
\end{tabular}
\end{table}

As shown in Table~\ref{tab:ablation_sequence}, we observe that:

1) \textbf{Hist-Only} performs worst in terms of online gain. While it captures stable user preferences, it is “static” within a single request cycle and fails to incorporate the fine-grained ranking signals from the user’s previous request, which reflect the system's real-time understanding of user intent.

2) \textbf{RankSig-Only} outperforms Hist-Only significantly in both offline and online metrics. This confirms that the partial-order of candidates contains rich, real-time "relative preference" information that is more critical for the retrieval-ranking alignment (Cascade Awareness).

3) \textbf{Fusion Trade-off}: RankSig-Only achieves the highest NDCG by closely mimicking downstream ranker preferences, yet POEM achieves superior HR@50 and HR@100. This indicates that fusing historical interactions anchors the user's intrinsic preferences while ranking signals steer the embedding toward the current candidate pool—broadening recall coverage beyond what a pure ranker-mimicker can achieve.

\subsubsection{Effectiveness of Hierarchical Learning Objective}
We investigate the components of our hierarchical learning objective, focusing on the synergy of positive sampling and the impact of hard negative mining.

\begin{table}[h]
\centering
\caption{Ablation study on positive sampling strategies. \texttt{Only-FB} uses only user feedback, while \texttt{Dual-Pos} (POEM) incorporates both user feedback and system-preferred items.}
\label{tab:ablation_positive}
\begin{tabular}{lcccccccc}
\toprule
\textbf{Strategy}  & \textbf{HR@50} & \textbf{NDCG@50} & \textbf{\makecell{Online\\ Gain}} & \textbf{\makecell{Delivery \\Rate}} \\
\midrule
Only-FB  & {0.5661} & {0.2434}  & +0.161\% & 1.052\% \\
Only-Sys  & 0.5966 & {0.2637} & +0.132\% & 1.472\% \\
w/o Hard-Neg & 0.6154 & 0.2774 & - & - \\
\midrule
\textbf{\makecell{Dual-Pos\\ (POEM)}} & \textbf{0.6661}  & \textbf{0.3079} & \textbf{+0.254\%} & \textbf{1.845\%} \\
\bottomrule
\end{tabular}
\end{table}

\paragraph{Analysis of Positive Sampling Strategies} As illustrated in Table~\ref{tab:ablation_positive}, the \texttt{Dual-Pos} strategy (POEM) significantly outperforms single-sampling variants in both offline alignment and online utility. We observe that neither user feedback nor system preference alone is sufficient:

\begin{itemize}[leftmargin=*]
\item \textbf{Limitations of \texttt{Only-FB}}: While $i^+_{\text{fb}}$ represents genuine user satisfaction, it suffers from severe \textbf{Selection Bias} and sparsity. Since $i^+_{\text{fb}}$ is derived only from items already exposed and clicked, training solely on it forces the retrieval model to overfit the historical exposure policy. This leads to \textbf{Objective Misalignment}: retrieval may capture items with high click probability but low multi-objective utility (e.g., short watch time), which are subsequently filtered out by downstream rankers. This is evidenced by the lowest \textbf{Delivery Rate} (1.052\%), indicating poor alignment with the overall system goal.

\item \textbf{Role of \texttt{Only-Sys}}: In contrast, incorporating $i^+_{\text{sys}}$ provides \textbf{Cascade Awareness}. By learning from the system's top-ranked items, the model "distills" the multi-objective knowledge of the ranking stage. Although this slightly shifts the focus from raw clicks, it significantly improves \textbf{Delivery Efficiency}. As shown, \texttt{Only-Sys} already achieves a higher Delivery Rate (1.472\%) and Online Gain than \texttt{Only-FB}, proving that aligning with the ranker is crucial for the survival of candidates in the pipeline.

\end{itemize}

\paragraph{Impact of Hard Negative Mining.} 
Next, we evaluate the contribution of dynamic graph-based hard negative mining. As shown in Table~\ref{tab:ablation_positive}, replacing our hard negatives with simple random negatives (\textbf{w/o Hard-Neg}) leads to a substantial drop in HR@50 (from 0.6661 to 0.6154) and NDCG@50. This confirms that in global-scale systems, random negatives are often too "easy" to provide sufficient gradient information. By mining semantically related but non-interacted items via the Swing graph, POEM learns a more precise decision boundary, effectively distinguishing subtle user preferences in a crowded candidate space.

\subsubsection{Signal Fusion Mechanism}

We evaluate five fusion strategies over three ranking signals—\textbf{CTR} (effective view, watch time $> A$), \textbf{CVR} (long view, watch time $> B > A$), and \textbf{WT} (raw watch time)—ranging from single-signal baselines to averaging and our rank-weight fusion (Multi-Rank). As shown in Table~\ref{tab:ablation_fusion}, Multi-Rank outperforms all variants.

Simple averaging (Multi-Avg) fails to effectively combine different signals, even underperforming compared to using a strong single signal (e.g., Single-WT). This demonstrates the necessity of handling scale misalignment and varying objective importance through our rank-weight based fusion method.

\begin{table}[h]
\centering
\caption{Ablation on signal fusion mechanisms.}
\label{tab:ablation_fusion}
\begin{tabular}{lcccc}
\toprule
\textbf{Variant} & \textbf{\makecell{HR\\@50}} & \textbf{\makecell{NDCG\\@50}} & \textbf{\makecell{HR\\@100}} & \textbf{\makecell{NDCG\\@100}} \\
\midrule
Single-CTR & 0.5690  & 0.2452 & 0.6891 & 0.2647 \\
Single-CVR & 0.5579 & 0.1781 & 0.6796 & 0.2348 \\
Single-WT & 0.5820 & 0.2507 & 0.6987 & 0.2697 \\
Multi-Avg & 0.5224 & 0.2257 & 0.6353 & 0.2440 \\
\midrule
\textbf{\makecell{Multi-Rank(POEM)}} & \textbf{0.6661}  & \textbf{0.3079} & \textbf{0.7725} & \textbf{0.3253} \\
\bottomrule
\end{tabular}
\end{table}

\subsection{Case Studies}
\label{subsec:case}

\begin{figure}[htbp]
    \centering
    \includegraphics[width=0.99\linewidth]{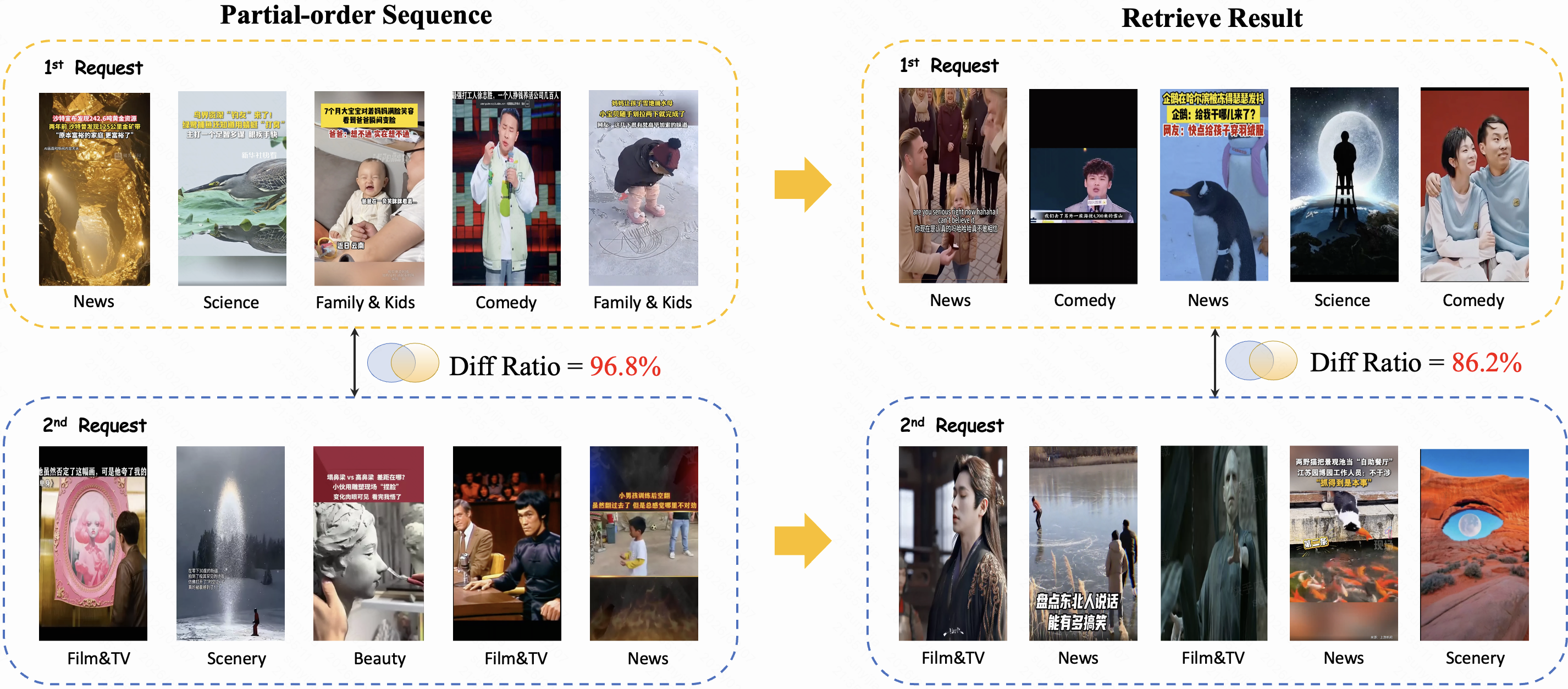}
    \caption{Case study showing the evolution of input sequences and retrieval results. (Left) Top items in Partial-order Sequence $S_r$ change as ranking scores update. (Right) Corresponding shifts in the final top retrieval results, demonstrating real-time interest adaptation.}
    \label{fig:case_diff}
\end{figure}

To quantify POEM's real-time responsiveness, we visualize the constructed partial-order sequences for a typical user across two consecutive refresh requests. As shown in Figure \ref{fig:case_diff}, although no new user clicks occurred during this interval, the input partial-order sequence $S_r$ exhibits a sequence diff ratio of 96.8\%. 
This significant change is driven by the interplay between the evolving candidate pool and our adaptive sampling strategy, which captures the latest ranking signals from the preceding stage.

This shift in $S_r$ effectively "steers" the user embedding $\mathbf{h}_u$ in the vector space, leading to an 86.2\% difference in the final retrieval results. As visualized by the item screenshots, the system captures a subtle transition from "Comedy" to "Film \& TV" based on the latest candidate availability. This demonstrates that POEM can adapt to the system's latest understanding of the user even during "passive" periods, whereas static sequence models (e.g., SASRec) would yield identical results.

\section{Conclusion}
\label{sec:conclusion}

In this paper, we introduced \textbf{POEM (Partial-Order Enhanced Modeling)}, a novel real-time recommendation framework that directly integrates cascade intelligence into sequential user modeling. To bridge the gap between static historical sequences and dynamic system context, POEM constructs a \textit{partial-order sequence} from the multi-objective ranking signals of the user's preceding request, enabling request-level re-evaluation of user interest. POEM achieves this through a dual-sequence architecture, a group-random sampling strategy for partial-order construction, and a hierarchical objective with hard negative mining.

Extensive evaluations on large-scale Kuaishou data validate POEM's effectiveness. Offline, it outperforms SASRec and CAIN in accuracy. More importantly, online A/B tests across two major scenarios confirm its practical impact, with significant gains in user engagement (\textbf{+0.249\%} and \textbf{+0.213\%} in average usage time per user). These results substantiate that leveraging ranking signals as a structural prior is key to real-time, cascade-aware recommendation. Future work includes improving robustness to noisy ranking signals and extending the partial-order paradigm to cross-domain settings, where system intelligence can guide broader exploration.

\bibliographystyle{ACM-Reference-Format}
\bibliography{sample-base}


\begin{thebibliography}{35}

%%% the Web address.

\ifx \showCODEN    \undefined \def \showCODEN     #1{\unskip}     \fi
\ifx \showISBNx    \undefined \def \showISBNx     #1{\unskip}     \fi
\ifx \showISBNxiii \undefined \def \showISBNxiii  #1{\unskip}     \fi
\ifx \showISSN     \undefined \def \showISSN      #1{\unskip}     \fi
\ifx \showLCCN     \undefined \def \showLCCN      #1{\unskip}     \fi
\ifx \shownote     \undefined \def \shownote      #1{#1}          \fi
\ifx \showarticletitle \undefined \def \showarticletitle #1{#1}   \fi
\ifx \showURL      \undefined \def \showURL       {\relax}        \fi
% invisible to TeX
\providecommand\bibfield[2]{#2}
\providecommand\bibinfo[2]{#2}
\providecommand\natexlab[1]{#1}
\providecommand\showeprint[2][]{arXiv:#2}

\bibitem[Bruch et~al\mbox{.}(2020)]%
        {bruch2020stochastic}
\bibfield{author}{\bibinfo{person}{Sebastian Bruch}, \bibinfo{person}{Shuguang Han}, \bibinfo{person}{Michael Bendersky}, {and} \bibinfo{person}{Marc Najork}.} \bibinfo{year}{2020}\natexlab{}.
\newblock \showarticletitle{A stochastic treatment of learning to rank scoring functions}. In \bibinfo{booktitle}{\emph{Proceedings of the 13th international conference on web search and data mining}}. \bibinfo{pages}{61--69}.
\newblock


\bibitem[Burges(2010)]%
        {burges2010ranknet}
\bibfield{author}{\bibinfo{person}{Christopher~JC Burges}.} \bibinfo{year}{2010}\natexlab{}.
\newblock \showarticletitle{From ranknet to lambdarank to lambdamart: An overview}.
\newblock \bibinfo{journal}{\emph{Learning}} \bibinfo{volume}{11}, \bibinfo{number}{23-581} (\bibinfo{year}{2010}), \bibinfo{pages}{81}.
\newblock


\bibitem[Cao et~al\mbox{.}(2025)]%
        {cao2025pantheon}
\bibfield{author}{\bibinfo{person}{Jiangxia Cao}, \bibinfo{person}{Pengbo Xu}, \bibinfo{person}{Yin Cheng}, \bibinfo{person}{Kaiwei Guo}, \bibinfo{person}{Jian Tang}, \bibinfo{person}{Shijun Wang}, \bibinfo{person}{Dewei Leng}, \bibinfo{person}{Shuang Yang}, \bibinfo{person}{Zhaojie Liu}, \bibinfo{person}{Yanan Niu}, {et~al\mbox{.}}} \bibinfo{year}{2025}\natexlab{}.
\newblock \showarticletitle{Pantheon: Personalized multi-objective ensemble sort via iterative pareto policy optimization}. In \bibinfo{booktitle}{\emph{Proceedings of the 34th ACM International Conference on Information and Knowledge Management}}. \bibinfo{pages}{5575--5582}.
\newblock


\bibitem[Guo et~al\mbox{.}(2017)]%
        {guo2017deepfm}
\bibfield{author}{\bibinfo{person}{Huifeng Guo}, \bibinfo{person}{Ruiming Tang}, \bibinfo{person}{Yunming Ye}, \bibinfo{person}{Zhenguo Li}, {and} \bibinfo{person}{Xiuqiang He}.} \bibinfo{year}{2017}\natexlab{}.
\newblock \bibinfo{title}{DeepFM: A Factorization-Machine based Neural Network for CTR Prediction}.
\newblock
\showeprint[arxiv]{1703.04247}~[cs.IR]
\urldef\tempurl%
\url{https://arxiv.org/abs/1703.04247}
\showURL{%
\tempurl}


\bibitem[Guo et~al\mbox{.}(2025)]%
        {guo2025context}
\bibfield{author}{\bibinfo{person}{Ting Guo}, \bibinfo{person}{Zhaoyang Yang}, \bibinfo{person}{Qinsong Zeng}, {and} \bibinfo{person}{Ming Chen}.} \bibinfo{year}{2025}\natexlab{}.
\newblock \showarticletitle{Context-Aware Lifelong Sequential Modeling for Online Click-Through Rate Prediction}.
\newblock \bibinfo{journal}{\emph{arXiv preprint arXiv:2502.12634}} (\bibinfo{year}{2025}).
\newblock


\bibitem[Hidasi et~al\mbox{.}(2015)]%
        {hidasi2015session}
\bibfield{author}{\bibinfo{person}{Bal{\'a}zs Hidasi}, \bibinfo{person}{Alexandros Karatzoglou}, \bibinfo{person}{Linas Baltrunas}, {and} \bibinfo{person}{Domonkos Tikk}.} \bibinfo{year}{2015}\natexlab{}.
\newblock \showarticletitle{Session-based recommendations with recurrent neural networks}.
\newblock \bibinfo{journal}{\emph{arXiv preprint arXiv:1511.06939}} (\bibinfo{year}{2015}).
\newblock


\bibitem[Huang et~al\mbox{.}(2013)]%
        {huang2013dssm}
\bibfield{author}{\bibinfo{person}{Po-Sen Huang}, \bibinfo{person}{Xiaodong He}, \bibinfo{person}{Jianfeng Gao}, \bibinfo{person}{Li Deng}, \bibinfo{person}{Alex Acero}, {and} \bibinfo{person}{Larry Heck}.} \bibinfo{year}{2013}\natexlab{}.
\newblock \showarticletitle{Learning deep structured semantic models for web search using clickthrough data}. In \bibinfo{booktitle}{\emph{Proceedings of the 22nd ACM international conference on Information \& Knowledge Management}}. \bibinfo{pages}{2333--2338}.
\newblock


\bibitem[Kang et~al\mbox{.}(2023)]%
        {kang2023distillation}
\bibfield{author}{\bibinfo{person}{SeongKu Kang}, \bibinfo{person}{Wonbin Kweon}, \bibinfo{person}{Dongha Lee}, \bibinfo{person}{Jianxun Lian}, \bibinfo{person}{Xing Xie}, {and} \bibinfo{person}{Hwanjo Yu}.} \bibinfo{year}{2023}\natexlab{}.
\newblock \showarticletitle{Distillation from heterogeneous models for top-K recommendation}. In \bibinfo{booktitle}{\emph{Proceedings of the ACM Web Conference 2023}}. \bibinfo{pages}{801--811}.
\newblock


\bibitem[Kang and McAuley(2018)]%
        {kang2018self}
\bibfield{author}{\bibinfo{person}{Wang-Cheng Kang} {and} \bibinfo{person}{Julian McAuley}.} \bibinfo{year}{2018}\natexlab{}.
\newblock \showarticletitle{Self-attentive sequential recommendation}. In \bibinfo{booktitle}{\emph{2018 IEEE international conference on data mining (ICDM)}}. IEEE, \bibinfo{pages}{197--206}.
\newblock


\bibitem[Khani et~al\mbox{.}(2024)]%
        {khani2024bridging}
\bibfield{author}{\bibinfo{person}{Nikhil Khani}, \bibinfo{person}{Li Wei}, \bibinfo{person}{Aniruddh Nath}, \bibinfo{person}{Shawn Andrews}, \bibinfo{person}{Shuo Yang}, \bibinfo{person}{Yang Liu}, \bibinfo{person}{Pendo Abbo}, \bibinfo{person}{Maciej Kula}, \bibinfo{person}{Jarrod Kahn}, \bibinfo{person}{Zhe Zhao}, {et~al\mbox{.}}} \bibinfo{year}{2024}\natexlab{}.
\newblock \showarticletitle{Bridging the gap: Unpacking the hidden challenges in knowledge distillation for online ranking systems}. In \bibinfo{booktitle}{\emph{Proceedings of the 18th ACM Conference on Recommender Systems}}. \bibinfo{pages}{758--761}.
\newblock


\bibitem[Koren et~al\mbox{.}(2009)]%
        {koren2009matrix}
\bibfield{author}{\bibinfo{person}{Yehuda Koren}, \bibinfo{person}{Robert Bell}, {and} \bibinfo{person}{Chris Volinsky}.} \bibinfo{year}{2009}\natexlab{}.
\newblock \showarticletitle{Matrix factorization techniques for recommender systems}.
\newblock \bibinfo{journal}{\emph{Computer}} \bibinfo{volume}{42}, \bibinfo{number}{8} (\bibinfo{year}{2009}), \bibinfo{pages}{30--37}.
\newblock


\bibitem[Li et~al\mbox{.}(2020)]%
        {li2020time}
\bibfield{author}{\bibinfo{person}{Jiacheng Li}, \bibinfo{person}{Yujie Wang}, {and} \bibinfo{person}{Julian McAuley}.} \bibinfo{year}{2020}\natexlab{}.
\newblock \showarticletitle{Time interval aware self-attention for sequential recommendation}. In \bibinfo{booktitle}{\emph{Proceedings of the 13th international conference on web search and data mining}}. \bibinfo{pages}{322--330}.
\newblock


\bibitem[Li et~al\mbox{.}(2025)]%
        {liu2025mgsd}
\bibfield{author}{\bibinfo{person}{Liang Li}, \bibinfo{person}{Zhou Yang}, {and} \bibinfo{person}{Xiaofei Zhu}.} \bibinfo{year}{2025}\natexlab{}.
\newblock \showarticletitle{Multi-Granularity Sequence Denoising with Weakly Supervised Signal for Sequential Recommendation}.
\newblock \bibinfo{journal}{\emph{arXiv preprint arXiv:2510.10564}} (\bibinfo{year}{2025}).
\newblock


\bibitem[Lian et~al\mbox{.}(2018)]%
        {lian2018xdeepfm}
\bibfield{author}{\bibinfo{person}{Jianxun Lian}, \bibinfo{person}{Xiaohuan Zhou}, \bibinfo{person}{Fuzheng Zhang}, \bibinfo{person}{Zhongxia Chen}, \bibinfo{person}{Xing Xie}, {and} \bibinfo{person}{Guangzhong Sun}.} \bibinfo{year}{2018}\natexlab{}.
\newblock \showarticletitle{xdeepfm: Combining explicit and implicit feature interactions for recommender systems}. In \bibinfo{booktitle}{\emph{Proceedings of the 24th ACM SIGKDD international conference on knowledge discovery \& data mining}}. \bibinfo{pages}{1754--1763}.
\newblock


\bibitem[Liu et~al\mbox{.}(2025)]%
        {liu2025bidding}
\bibfield{author}{\bibinfo{person}{Bin Liu}, \bibinfo{person}{Yunfei Liu}, \bibinfo{person}{Ziru Xu}, \bibinfo{person}{Zhaoyu Zhou}, \bibinfo{person}{Zhi Kou}, \bibinfo{person}{Yeqiu Yang}, \bibinfo{person}{Han Zhu}, \bibinfo{person}{Jian Xu}, {and} \bibinfo{person}{Bo Zheng}.} \bibinfo{year}{2025}\natexlab{}.
\newblock \showarticletitle{Bidding-Aware Retrieval for Multi-Stage Consistency in Online Advertising}.
\newblock \bibinfo{journal}{\emph{arXiv preprint arXiv:2508.05206}} (\bibinfo{year}{2025}).
\newblock


\bibitem[Lu et~al\mbox{.}(2025)]%
        {lu2025mimn}
\bibfield{author}{\bibinfo{person}{Hui Lu}, \bibinfo{person}{Zheng Chai}, \bibinfo{person}{Yuchao Zheng}, \bibinfo{person}{Zhe Chen}, \bibinfo{person}{Deping Xie}, \bibinfo{person}{Peng Xu}, \bibinfo{person}{Xun Zhou}, {and} \bibinfo{person}{Di Wu}.} \bibinfo{year}{2025}\natexlab{}.
\newblock \showarticletitle{Large Memory Network for Recommendation}. In \bibinfo{booktitle}{\emph{Companion Proceedings of the ACM on Web Conference 2025}}. \bibinfo{pages}{1162--1166}.
\newblock


\bibitem[Ma et~al\mbox{.}(2018)]%
        {ma2018entire}
\bibfield{author}{\bibinfo{person}{Xiao Ma}, \bibinfo{person}{Liqin Zhao}, \bibinfo{person}{Guan Huang}, \bibinfo{person}{Zhi Wang}, \bibinfo{person}{Zelin Hu}, \bibinfo{person}{Xiaoqiang Zhu}, {and} \bibinfo{person}{Kun Gai}.} \bibinfo{year}{2018}\natexlab{}.
\newblock \showarticletitle{Entire space multi-task model: An effective approach for estimating post-click conversion rate}. In \bibinfo{booktitle}{\emph{The 41st International ACM SIGIR Conference on Research \& Development in Information Retrieval}}. \bibinfo{pages}{1137--1140}.
\newblock


\bibitem[Mao et~al\mbox{.}(2024)]%
        {mao2024analysis}
\bibfield{author}{\bibinfo{person}{Chunyan Mao}, \bibinfo{person}{Shuaishuai Huang}, \bibinfo{person}{Mingxiu Sui}, \bibinfo{person}{Haowei Yang}, {and} \bibinfo{person}{Xueshe Wang}.} \bibinfo{year}{2024}\natexlab{}.
\newblock \showarticletitle{Analysis and design of a personalized recommendation system based on a dynamic user interest model}.
\newblock \bibinfo{journal}{\emph{arXiv preprint arXiv:2410.09923}} (\bibinfo{year}{2024}).
\newblock


\bibitem[Pang et~al\mbox{.}(2020)]%
        {pang2020setrank}
\bibfield{author}{\bibinfo{person}{Liang Pang}, \bibinfo{person}{Jun Xu}, \bibinfo{person}{Qingyao Ai}, \bibinfo{person}{Yanyan Lan}, \bibinfo{person}{Xueqi Cheng}, {and} \bibinfo{person}{Jirong Wen}.} \bibinfo{year}{2020}\natexlab{}.
\newblock \showarticletitle{Setrank: Learning a permutation-invariant ranking model for information retrieval}. In \bibinfo{booktitle}{\emph{Proceedings of the 43rd international ACM SIGIR conference on research and development in information retrieval}}. \bibinfo{pages}{499--508}.
\newblock


\bibitem[Pi et~al\mbox{.}(2020)]%
        {pi2020search}
\bibfield{author}{\bibinfo{person}{Qi Pi}, \bibinfo{person}{Guorui Zhou}, \bibinfo{person}{Yujing Zhang}, \bibinfo{person}{Zhe Wang}, \bibinfo{person}{Lejian Ren}, \bibinfo{person}{Ying Fan}, \bibinfo{person}{Xiaoqiang Zhu}, {and} \bibinfo{person}{Kun Gai}.} \bibinfo{year}{2020}\natexlab{}.
\newblock \showarticletitle{Search-based user interest modeling with lifelong sequential behavior data for click-through rate prediction}. In \bibinfo{booktitle}{\emph{Proceedings of the 29th ACM International Conference on Information \& Knowledge Management}}. \bibinfo{pages}{2685--2692}.
\newblock


\bibitem[Rendle et~al\mbox{.}(2010)]%
        {rendle2010factorizing}
\bibfield{author}{\bibinfo{person}{Steffen Rendle}, \bibinfo{person}{Christoph Freudenthaler}, {and} \bibinfo{person}{Lars Schmidt-Thieme}.} \bibinfo{year}{2010}\natexlab{}.
\newblock \showarticletitle{Factorizing personalized markov chains for next-basket recommendation}. In \bibinfo{booktitle}{\emph{Proceedings of the 19th international conference on World wide web}}. \bibinfo{pages}{811--820}.
\newblock


\bibitem[Sun et~al\mbox{.}(2019)]%
        {sun2019bert4rec}
\bibfield{author}{\bibinfo{person}{Fei Sun}, \bibinfo{person}{Jun Liu}, \bibinfo{person}{Jian Wu}, \bibinfo{person}{Changhua Pei}, \bibinfo{person}{Xiao Lin}, \bibinfo{person}{Wenwu Ou}, {and} \bibinfo{person}{Peng Jiang}.} \bibinfo{year}{2019}\natexlab{}.
\newblock \showarticletitle{BERT4Rec: Sequential recommendation with bidirectional encoder representations from transformer}. In \bibinfo{booktitle}{\emph{Proceedings of the 28th ACM international conference on information and knowledge management}}. \bibinfo{pages}{1441--1450}.
\newblock


\bibitem[Sun et~al\mbox{.}(2025a)]%
        {sun2025mpformer}
\bibfield{author}{\bibinfo{person}{Yijia Sun}, \bibinfo{person}{Shanshan Huang}, \bibinfo{person}{Linxiao Che}, \bibinfo{person}{Haitao Lu}, \bibinfo{person}{Qiang Luo}, \bibinfo{person}{Kun Gai}, {and} \bibinfo{person}{Guorui Zhou}.} \bibinfo{year}{2025}\natexlab{a}.
\newblock \showarticletitle{MPFormer: Adaptive Framework for Industrial Multi-Task Personalized Sequential Retriever}. In \bibinfo{booktitle}{\emph{Proceedings of the 34th ACM International Conference on Information and Knowledge Management}}. \bibinfo{pages}{2832--2841}.
\newblock


\bibitem[Sun et~al\mbox{.}(2025b)]%
        {sun2025grank}
\bibfield{author}{\bibinfo{person}{Yijia Sun}, \bibinfo{person}{Shanshan Huang}, \bibinfo{person}{Zhiyuan Guan}, \bibinfo{person}{Qiang Luo}, \bibinfo{person}{Ruiming Tang}, \bibinfo{person}{Kun Gai}, {and} \bibinfo{person}{Guorui Zhou}.} \bibinfo{year}{2025}\natexlab{b}.
\newblock \showarticletitle{GRank: Towards Target-Aware and Streamlined Industrial Retrieval with a Generate-Rank Framework}.
\newblock \bibinfo{journal}{\emph{arXiv preprint arXiv:2510.15299}} (\bibinfo{year}{2025}).
\newblock


\bibitem[Tang and Wang(2018a)]%
        {tang2018caser}
\bibfield{author}{\bibinfo{person}{Jiaxi Tang} {and} \bibinfo{person}{Ke Wang}.} \bibinfo{year}{2018}\natexlab{a}.
\newblock \showarticletitle{Personalized top-n sequential recommendation via convolutional sequence embedding}. In \bibinfo{booktitle}{\emph{Proceedings of the eleventh ACM international conference on web search and data mining}}. \bibinfo{pages}{565--573}.
\newblock


\bibitem[Tang and Wang(2018b)]%
        {tang2018ranking}
\bibfield{author}{\bibinfo{person}{Jiaxi Tang} {and} \bibinfo{person}{Ke Wang}.} \bibinfo{year}{2018}\natexlab{b}.
\newblock \showarticletitle{Ranking distillation: Learning compact ranking models with high performance for recommender system}. In \bibinfo{booktitle}{\emph{Proceedings of the 24th ACM SIGKDD international conference on knowledge discovery \& data mining}}. \bibinfo{pages}{2289--2298}.
\newblock


\bibitem[Vaswani et~al\mbox{.}(2017)]%
        {vaswani2017attention}
\bibfield{author}{\bibinfo{person}{Ashish Vaswani}, \bibinfo{person}{Noam Shazeer}, \bibinfo{person}{Niki Parmar}, \bibinfo{person}{Jakob Uszkoreit}, \bibinfo{person}{Llion Jones}, \bibinfo{person}{Aidan~N Gomez}, \bibinfo{person}{{\L}ukasz Kaiser}, {and} \bibinfo{person}{Illia Polosukhin}.} \bibinfo{year}{2017}\natexlab{}.
\newblock \showarticletitle{Attention is all you need}.
\newblock \bibinfo{journal}{\emph{Advances in neural information processing systems}}  \bibinfo{volume}{30} (\bibinfo{year}{2017}).
\newblock


\bibitem[Wilm et~al\mbox{.}(2023)]%
        {wilm2023scaling}
\bibfield{author}{\bibinfo{person}{Timo Wilm}, \bibinfo{person}{Philipp Normann}, \bibinfo{person}{Sophie Baumeister}, {and} \bibinfo{person}{Paul-Vincent Kobow}.} \bibinfo{year}{2023}\natexlab{}.
\newblock \showarticletitle{Scaling session-based transformer recommendations using optimized negative sampling and loss functions}. In \bibinfo{booktitle}{\emph{Proceedings of the 17th ACM conference on recommender systems}}. \bibinfo{pages}{1023--1026}.
\newblock


\bibitem[Wu et~al\mbox{.}(2025)]%
        {wu2025rankcot}
\bibfield{author}{\bibinfo{person}{Mingyan Wu}, \bibinfo{person}{Zhenghao Liu}, \bibinfo{person}{Yukun Yan}, \bibinfo{person}{Xinze Li}, \bibinfo{person}{Shi Yu}, \bibinfo{person}{Zheni Zeng}, \bibinfo{person}{Yu Gu}, {and} \bibinfo{person}{Ge Yu}.} \bibinfo{year}{2025}\natexlab{}.
\newblock \showarticletitle{RankCoT: Refining Knowledge for Retrieval-Augmented Generation through Ranking Chain-of-Thoughts}.
\newblock \bibinfo{journal}{\emph{arXiv preprint arXiv:2502.17888}} (\bibinfo{year}{2025}).
\newblock


\bibitem[Xie et~al\mbox{.}(2022)]%
        {2022cl4srec}
\bibfield{author}{\bibinfo{person}{Xu Xie}, \bibinfo{person}{Fei Sun}, \bibinfo{person}{Zhaoyang Liu}, \bibinfo{person}{Shiwen Wu}, \bibinfo{person}{Jinyang Gao}, \bibinfo{person}{Jiandong Zhang}, \bibinfo{person}{Bolin Ding}, {and} \bibinfo{person}{Bin Cui}.} \bibinfo{year}{2022}\natexlab{}.
\newblock \showarticletitle{Contrastive learning for sequential recommendation}. In \bibinfo{booktitle}{\emph{2022 IEEE 38th international conference on data engineering (ICDE)}}. IEEE, \bibinfo{pages}{1259--1273}.
\newblock


\bibitem[Yang et~al\mbox{.}(2020)]%
        {yang2020large}
\bibfield{author}{\bibinfo{person}{Xiaoyong Yang}, \bibinfo{person}{Yadong Zhu}, \bibinfo{person}{Yi Zhang}, \bibinfo{person}{Xiaobo Wang}, {and} \bibinfo{person}{Quan Yuan}.} \bibinfo{year}{2020}\natexlab{}.
\newblock \showarticletitle{Large scale product graph construction for recommendation in e-commerce}.
\newblock \bibinfo{journal}{\emph{arXiv preprint arXiv:2010.05525}} (\bibinfo{year}{2020}).
\newblock


\bibitem[Yi et~al\mbox{.}(2019)]%
        {yi2019sampling}
\bibfield{author}{\bibinfo{person}{Xinyang Yi}, \bibinfo{person}{Ji Yang}, \bibinfo{person}{Lichan Hong}, \bibinfo{person}{Derek~Zhiyuan Cheng}, \bibinfo{person}{Lukasz Heldt}, \bibinfo{person}{Aditee Kumthekar}, \bibinfo{person}{Zhe Zhao}, \bibinfo{person}{Li Wei}, {and} \bibinfo{person}{Ed Chi}.} \bibinfo{year}{2019}\natexlab{}.
\newblock \showarticletitle{Sampling-bias-corrected neural modeling for large corpus item recommendations}. In \bibinfo{booktitle}{\emph{Proceedings of the 13th ACM conference on recommender systems}}. \bibinfo{pages}{269--277}.
\newblock


\bibitem[Zhao et~al\mbox{.}(2023)]%
        {zhao2023copr}
\bibfield{author}{\bibinfo{person}{Zhishan Zhao}, \bibinfo{person}{Jingyue Gao}, \bibinfo{person}{Yu Zhang}, \bibinfo{person}{Shuguang Han}, \bibinfo{person}{Siyuan Lou}, \bibinfo{person}{Xiang-Rong Sheng}, \bibinfo{person}{Zhe Wang}, \bibinfo{person}{Han Zhu}, \bibinfo{person}{Yuning Jiang}, \bibinfo{person}{Jian Xu}, {et~al\mbox{.}}} \bibinfo{year}{2023}\natexlab{}.
\newblock \showarticletitle{Copr: Consistency-oriented pre-ranking for online advertising}. In \bibinfo{booktitle}{\emph{Proceedings of the 32nd ACM International Conference on Information and Knowledge Management}}. \bibinfo{pages}{4974--4980}.
\newblock


\bibitem[Zhou et~al\mbox{.}(2019)]%
        {zhou2019deep}
\bibfield{author}{\bibinfo{person}{Guorui Zhou}, \bibinfo{person}{Na Mou}, \bibinfo{person}{Ying Fan}, \bibinfo{person}{Qi Pi}, \bibinfo{person}{Weijie Bian}, \bibinfo{person}{Chang Zhou}, \bibinfo{person}{Xiaoqiang Zhu}, {and} \bibinfo{person}{Kun Gai}.} \bibinfo{year}{2019}\natexlab{}.
\newblock \showarticletitle{Deep interest evolution network for click-through rate prediction}. In \bibinfo{booktitle}{\emph{Proceedings of the AAAI conference on artificial intelligence}}, Vol.~\bibinfo{volume}{33}. \bibinfo{pages}{5941--5948}.
\newblock


\bibitem[Zhou et~al\mbox{.}(2018)]%
        {zhou2018deep}
\bibfield{author}{\bibinfo{person}{Guorui Zhou}, \bibinfo{person}{Xiaoqiang Zhu}, \bibinfo{person}{Chenru Song}, \bibinfo{person}{Ying Fan}, \bibinfo{person}{Han Zhu}, \bibinfo{person}{Xiao Ma}, \bibinfo{person}{Yanghui Yan}, \bibinfo{person}{Junqi Jin}, \bibinfo{person}{Han Li}, {and} \bibinfo{person}{Kun Gai}.} \bibinfo{year}{2018}\natexlab{}.
\newblock \showarticletitle{Deep interest network for click-through rate prediction}. In \bibinfo{booktitle}{\emph{Proceedings of the 24th ACM SIGKDD international conference on knowledge discovery \& data mining}}. \bibinfo{pages}{1059--1068}.
\newblock


\end{thebibliography}

\end{document}